\documentclass[prb,preprint]{revtex4}
\usepackage{latexsym}
\usepackage{amsmath}
\usepackage{amssymb}
\usepackage{amsthm}
\usepackage{graphicx}
\usepackage{dcolumn}
\newcommand{\be}{\begin{equation}}
\newcommand{\ee}{\end{equation}}
\newcommand{\ba}{\begin{eqnarray}}
\newcommand{\ea}{\end{eqnarray}}

\begin{document}
\title{Anomalous melting behavior under extreme conditions: \\
hard matter turning ``soft''}
\author{Gianpietro Malescio$^1$~\cite{aff1}, Franz Saija$^2$~\cite{aff2},
and Santi Prestipino$^1$~\cite{aff3}}
\affiliation{$^1$ Universit\`a degli Studi di Messina, Dipartimento di Fisica,
Contrada Papardo, 98166 Messina, Italy \\
$^2$ CNR-Istituto per i Processi Chimico-Fisici, Contrada Papardo,
98158 Messina, Italy}
%\date{\today}
\begin{abstract}
We show that a system of particles interacting
through the exp-6 pair potential, commonly used to describe
effective interatomic forces under high compression,
exhibits anomalous melting features such as reentrant melting
and a rich solid polymorphism, including a stable BC8 crystal.
We relate this behavior to the crossover, with increasing pressure,
between two different regimes of local order that are associated with
the two repulsive length scales of the potential. Our results provide a
unifying picture for the high-pressure melting anomalies observed
in many elements and point out that, under extreme conditions,
atomic systems may reveal surprising similarities with soft matter.
\end{abstract}
\pacs{61.20.Ja, 62.50.-p, 64.70.D-, 64.70.K-}
\keywords{High-pressure phase diagrams of the elements, Liquid-solid
transitions, Solid-solid transitions, Reentrant melting}
\maketitle

Many atomic substances show anomalous melting features at high pressures,
where the term ``anomalous'' is commonly used to underline
any difference with respect to standard, simple-fluid-like
melting~\cite{McMillan,Young1,Saxena}.
However, some rare gases (usually assumed as prototypical simple
fluids) already deviate from this idealized behavior:
the slope ${\rm d}T/{\rm d}P$
of Ar, Kr and Xe melting lines shows a substantial decrease for high
pressures with respect
to predictions based on corresponding-state scaling from the Ne melting
curve~\cite{Boehler}. For other materials, the melting line $T_m(P)$ shows
an extended plateau (e.g. Ta, Mo, Cr~\cite{Errandonea1}) or even a maximum,
followed by a region of reentrant melting
(e.g. Cs, Rb, Na, Te, H, N~\cite{Jayaraman,Bundy,Gregoryanz,Raty,Errandonea2,Rapoport,Bonev,Mukherjee}).
In some cases (e.g. Cs, K, Ba, Na~\cite{Jayaraman,Zha,Errandonea2,Gregoryanz}),
a further pressure increase results in a positive melting slope again.
As advances in experimental methods allow to reach higher and
higher pressures, the class of substances exhibiting melting anomalies
constantly expands and materials that not long ago were thought to
undergo ``normal'' melting reveal instead an anomalous behavior
(e.g. Na~\cite{Gregoryanz}).

Melting features that are similar to those outlined above are known to occur
for classical pair interactions with a soft-repulsive
component~\cite{Royall,Stillinger,Likos,Prestipino,Malescio1,Malescio2,Stell,Young2,Jagla},
used for modeling colloidal suspensions and to study liquid-liquid
transitions and water-like anomalies. The crucial feature
of such interactions is the existence of a range of interparticle distances
where the strength of the repulsive force {\em reduces} as the distance gets
smaller (core-softening condition~\cite{Debenedetti1}). The
relevance of soft interactions for the high-pressure behavior of real
substances has received little attention so far since atoms are usually
thought to be ``hard'' objects. Yet, as will be argued below, extremely
high pressures may bring this common belief into question.
In order to explore the nature of melting at such high pressures,
we take a coarse-grained view of the problem and consider a classical
potential that is widely used to calculate the equation of state of
materials under extreme conditions, i.e., the Buckingham or exp-6
potential~\cite{Buckingham},
\be
u(r)=\left\{
\begin{array}{ll}
+\infty & ,\,\,\,r<\sigma_M \\
\frac{\epsilon}{\alpha-6}\left[
6\,e^{-\alpha\left(\frac{r}{\sigma}-1\right)}-
\alpha\left(\frac{\sigma}{r}\right)^6\right] & ,\,\,\,r\ge\sigma_M
\end{array}
\right.
\label{eq1}
\ee
where $r$ is the interparticle distance, $\epsilon$ is the depth of the
attractive well, $\sigma$ is the position of the well minimum, $\alpha$
(usually taken in the range 10-15~\cite{Fried}) controls the steepness
of the short-range repulsion, and $\sigma_M$ is the point where the function
in the second line of (1) attains its maximum value $\epsilon_M$.
It has been noted that the exp-6
model accounts for high-pressure effects in much better way than other
more popular models (e.g. the Lennard-Jones model) owing to its
less steep repulsion~\cite{Ross2}. However, the ability of the exp-6
interaction to satisfy the core-softening
condition (see Fig.\,1) has not been pointed out so far and its phase behavior
was investigated only in a restricted range of pressures and temperatures
where no melting anomaly occurs~\cite{Ross2,Belonoshko,Saija}.

Using standard simulation methods ($NPT$ Metropolis Monte Carlo in
conjunction with Widom and Frenkel-Ladd free-energy methods~\cite{Frenkel}),
we computed the exp-6 phase diagram for a given $\alpha$ (choosing
initially $\alpha=11$) over a wide range of
pressures $P$ and temperatures $T$ (expressed throughout the text in units
of $\epsilon/\sigma^3$ and $\epsilon/k_B$, respectively).
Our samples contained a number $N$ of particles of the order of 1000
(finite-size corrections are negligible).
As shown in Fig.\,2, the melting line starts at low $P$ with
a positive slope that gradually reduces with increasing pressure until
it vanishes at a point of maximum melting temperature $T_M$. This is
followed by a pressure interval where ${\rm d}T_m/{\rm d}P$ is negative.
Eventually, at extremely high pressures, $T_m(P)$ recovers a
positive slope. In order to discuss the phase diagram, assume for instance
to increase $P$ at fixed temperature (say $T=8$). The system,
initially fluid, becomes denser and denser with increasing pressure
until it crystallizes into a face-centered cubic (FCC) solid.
Upon increasing $P$ further, the FCC solid undergoes a transition into
a body-centered cubic (BCC) solid (FCC goes over into BCC also
upon heating at constant pressure). This effect is related to a
decrease in the mean nearest-neighbor distance with increasing pressure
(or temperature), which brings particles to experience inner, less steep
regions of the interaction potential (it is known that fluids interacting
via a inverse-power law, $u(r)\propto r^{-n}$, do crystallize into a
BCC solid for $n\lesssim 7$ while the solid phase is FCC for steeper
repulsions~\cite{Agrawal}). When $P$ is large enough, the BCC
solid finally melts into a denser fluid. This follows from increasing
competition, upon compression, between two separate scales of first-neighbor
distance, i.e., a larger one associated with the soft repulsion (effective
at the lower pressures) and a smaller one related to the
particle-core diameter $\sigma_M$ (dominant at the
higher pressures). These two characteristic lengths push for different
and incommensurate patterns of short-range order in the system, thus
setting the stage for reentrance of the fluid phase at intermediate
pressures (at least for not too low $T$).
Upon further compression, the smaller length scale eventually takes
over and the fluid crystallizes into a hard-sphere-like, FCC solid.

In the fluid region above the reentrant-melting line, isobaric cooling leads
first to an increase in the number density $\rho$, followed by a decrease
for further cooling (Fig.\,2, inset). The locus of points where $\rho$
attains its maximum encloses a region of ``density anomaly'' (see Fig.\,2).
This has been observed under ordinary pressure
in a number of systems, among which water is the most
important~\cite{Debenedetti2}. Our findings raise the possibility that a density
anomaly may as well occur at much higher pressures in those substances
that are characterized by reentrant melting.

The above results are confirmed and further corroborated by an analysis
of the system structure in terms of the radial distribution function $g(r)$
and of the structure factor $S(k)$. We computed $g(r)$ at
a temperature $T$ close to $T_M$ and in the pressure range where reentrant
melting occurs. Upon compression, the first peak of $g(r)$
(at $r\simeq\sigma_M$) moves upward while the second and third peaks go
down, signalling that more and more particles can overcome the soft-repulsive
shoulder (Fig.\,3, top panel). This behavior is mirrored in the pressure
dependence of $S(k)$ whose main peak first builds up and then goes down
(Fig.\,3, inset of top panel). This is different from simple fluids,
where all $g(r)$ and $S(k)$ peaks get higher as $P$ grows at constant $T$.
We also computed $g(r)$ at a somewhat lower temperature ($T=10$) and for
intermediate pressures. Here, $g(r)$ is highly structured for $r<\sigma$
while it shows an ideal-gas behavior for larger distances (Fig.\,3, bottom
panel). This suggests that, in the fluid region being considered, particles
are grouped in clusters whose linear extent is of the order of $\sigma$, with
inter-cluster spacing large enough that no significant radial correlations
exist between particles in different clusters. 

We did not attempt a full analysis of solid polymorphism in the exp-6
system at low temperature, but to a large extent this can be anticipated
by the $T=0$ calculation of the chemical potential (i.e., enthalpy)
for a number of relevant crystal structures (see Fig.\,4).
Optimized structures were computed for the following lattices:
FCC, BCC, hexagonal closed packed, simple cubic (SC),
simple hexagonal (SH), body-centered tetragonal, plus a few
non-Bravais lattices (diamond, BC8, cI16, hR1, and ST12) that occur,
either as stable or near-optimal phases, for Li, Na, and Si under high
compression~\cite{Neaton,Hanfland,McMahon,Tamblyn}.
We found that the sequence of stable phases for increasing pressures is
\be
{\rm FCC}\stackrel{5500}{\longrightarrow}{\rm BCC}
\stackrel{12500}{\longrightarrow}{\rm SH}
\stackrel{19500}{\longrightarrow}{\rm BC8}
\stackrel{31500}{\longrightarrow}{\rm SH}
\stackrel{42500}{\longrightarrow}{\rm SC}
\stackrel{53500}{\longrightarrow}{\rm SH}
\stackrel{72500}{\longrightarrow}{\rm FCC}\,,
\label{eq2}
\ee
where the numbers above the arrows indicate the transition pressures
(to within a precision of 500). The three separate SH regions correspond
to distinct ranges of the ratio $c/a$ between transverse and in-plane lattice
parameters, that is $0.69$--$0.73$, $1.33$--$1.23$,
and 1, respectively. In turn, the number of neighbors in close contact with
a given particle is 2, 6, and 8. Moreover, at exactly $P=20000$, the stable
BC8 crystal
can also be viewed as a cI16 crystal with an internal positional parameter
of 0.125. The occurrence of BC8 and cI16 solids (same Pearson
symbol~\cite{Pearson}) in
light alkali metals has been associated with complex modifications
of the electronic density of states~\cite{Neaton,Hanfland,Martin}.
Remarkably, our findings show that low-symmetry non-Bravais lattices
can be stabilized also for a simple spherically-symmetric classical
interaction.
The existence of two competing repulsive length scales appears to be
an essential ingredient for this surprising result, which discloses
the possibility that suitably-tailored colloids (i.e., micrometer-sized
particles) may exist as ``exotic'' solids at standard conditions.

We finally analysed how the phase diagram of the exp-6 model changes with
the repulsion steepness by performing calculations for other values
of $\alpha$ in the range 10-13. We found that the typical pressure and
temperature $(\widetilde{P},\widetilde{T})$ where anomalous features occur do
approximately scale as $\epsilon_M(\alpha)/\sigma_M(\alpha)^3$ and
$\epsilon_M(\alpha)$, respectively. This amounts to about a tenfold
($\widetilde{P}$) and a fivefold ($\widetilde{T}$) increase per unit $\alpha$
variation.
In addition to this major effect, we found that the maximum of $T_m(P)$
becomes more and more pronounced with increasing $\alpha$.

As illustrated above, the melting behavior of the exp-6 system is related
to the gradual switching off, under sufficiently high pressure, of the
soft-repulsive length scale in favor of the hard one, which leads to
a series of transitions to more and more compact structures. This offers
a clue to understand the anomalous melting features of many materials under
extreme conditions.
As is well known, pressure may trigger a reorganization of the atomic
structure, leading to charge transfer to more localized
orbitals (see e.g. the 6s-5d transition in Cs~\cite{Jayaraman}). A similar
phenomenon is pressure-driven 5p-5d hybridization in Xe~\cite{Ross1}.
More often, a pressure-induced symmetry-breaking transition
is the large-scale manifestation of a collective response of conduction
electrons: upon compression, pseudo-ions will eventually adjust to a new
and more compact crystal lattice provided this also ensures an optimum
electronic-energy content. Both atomic and structural transitions take place
at definite pressures in the solid while in the fluid such changes
occur over a broader pressure range. The transition of an element to a more
compact solid is usually reflected in a sudden increase of the
$T_m(P)$ slope. This is generally preceded by a part of the melting line
having negative, vanishing, or very small positive slope. Aside from the
specific mechanism at work, such behaviors can be interpreted, in the
effective-potential approach, as effects of the weakening of repulsive
forces that is associated with the crossover from a larger to a
smaller repulsive length scale. This affects, to a greater or lesser
extent, any substance and will induce its structure, at sufficiently high
pressure, to settle down on a more compact and stiff
arrangement. In some systems, this process may occur repeatedly as pressure
increases, which provides a rationale for the behavior
hypothesized for K~\cite{Zha} and observed in Sr~\cite{Errandonea2}.

The melting behavior that was for long
considered as general, i.e., a regularly increasing and concave $T_m(P)$
(typical of e.g. hard-sphere and inverse-power potentials), is actually
unrealistic at extreme pressures since it is associated with an essentially
rigid-like response to compression. On the contrary, anomalous
melting can be expected to be the norm among the elements.
However, the pressures and temperatures where structural softening occurs
can vary considerably from one substance to the other, as suggested by
the sensitive dependence on the repulsion steepness of the location of
exp-6 anomalies. In particular, atoms with more electrons
should be more susceptible, at least within the same chemical family,
to pressure-induced modifications in the condensed phases.
This is consistent, for example, with the known properties of alkali
metals~\cite{Jayaraman,Rapoport,Bundy,Zha,Gregoryanz,Raty} and with
the behavior of
rare gases, where the flattening observed in the melting line at
high pressures is more marked and occurs at a smaller pressure
for the heavier gases~\cite{Boehler,Ross1}.

\newpage
%
%  FIGURE CAPTIONS
%
\begin{center}
\large
FIGURE CAPTIONS
\normalsize
\end{center}
\begin{description}
\item[{\bf Fig.\,1 :}]
The exp-6 potential $u(r)$ for $\alpha=11$ (solid line, left vertical axis)
and the corresponding force $f(r)=-{\rm d}u/{\rm d}r$ (dashed line, right
vertical axis).
The force has a maximum strength in the region of the soft-repulsive core.
Hence, a new ``soft'' length scale emerges in the system in addition to
the hard-core diameter $\sigma_M$.
With increasing $\alpha$, $\sigma_M$ decreases while $\epsilon_M$ gets
larger. For instance, when $\alpha$ grows from 11 to 13, $\sigma_M$
varies from about $0.374\sigma$ to $0.245\sigma$ whereas $\epsilon_M$
varies from about $370\epsilon$ to $7104\epsilon$.

\item[{\bf Fig.\,2 :}]
High $P$-high $T$ phase diagram of the exp-6 model for $\alpha=11$.
Pressure $P$ and temperature $T$ are in units of $\epsilon/\sigma^3$
and $\epsilon/k_B$, respectively. Coexistence curves are represented
as solid lines. Open circles mark coexistence points as estimated
through exact free-energy calculations (errors are smaller than the
symbols size). The full circle is the outcome of a total-energy calculation.
The boundary between the fluid phase and the FCC crystal at extremely
high pressures corresponds to the lower stability threshold of the solid.
In the intermediate-pressure region (from about 15000 to 70000),
additional phases at low temperature are likely present (a BC8 solid,
various instances of SH solids, and a SC solid, see text), whose precise
boundaries, however, were not computed.
The locus of density maxima, $(\partial\rho/\partial T)_P=0$, in the fluid
phase is represented as a dotted line. All lines in the figure are guides
to the eye. Inset: reduced number density $\rho\sigma^3$ as a function of
temperature for $P=10000$.

\item[{\bf Fig.\,3 :}]
Spatial correlations in the exp-6 system for $\alpha=11$.
Top panel: radial distribution function $g(r)$ for $T=17$;
$P=9000$ (solid line), 12000 (dashed line), 15000 (dotted line), and 18000
(dash-dotted line). Inset: height of the first peak of the structure factor
$S(k)$ as a function of pressure at $T=17$.
Bottom panel: $g(r)$ for $T=10$; $P=15000$ (dotted line), 20000
(dash-dotted line), 40000 (dashed line), 60000 (solid line).

\item[{\bf Fig.\,4 :}]
(Color online) $T=0$ chemical potential $\mu$, plotted as a function of pressure $P$,
for a number of crystal arrangements, choosing the FCC lattice (blue solid
line) as a reference (both $\mu$ and $P$ are in reduced units; the chemical
potentials of structures that are never stable are not shown):
BCC (triangles and red solid line), simple cubic (SC, boxes and black
solid line), simple hexagonal (SH, dots and light blue solid line),
BC8 (magenta dotted line), and cI16 (green dashed line). The optimized cI16
crystal is actually a BCC crystal up to the pressure where the BC8 solid
prevails over BCC. From there onwards, the cI16 line runs over the BC8 line,
departing from it only for $P>20000$.
\end{description}

\newpage
\begin{figure}
\includegraphics[width=12cm,angle=0]{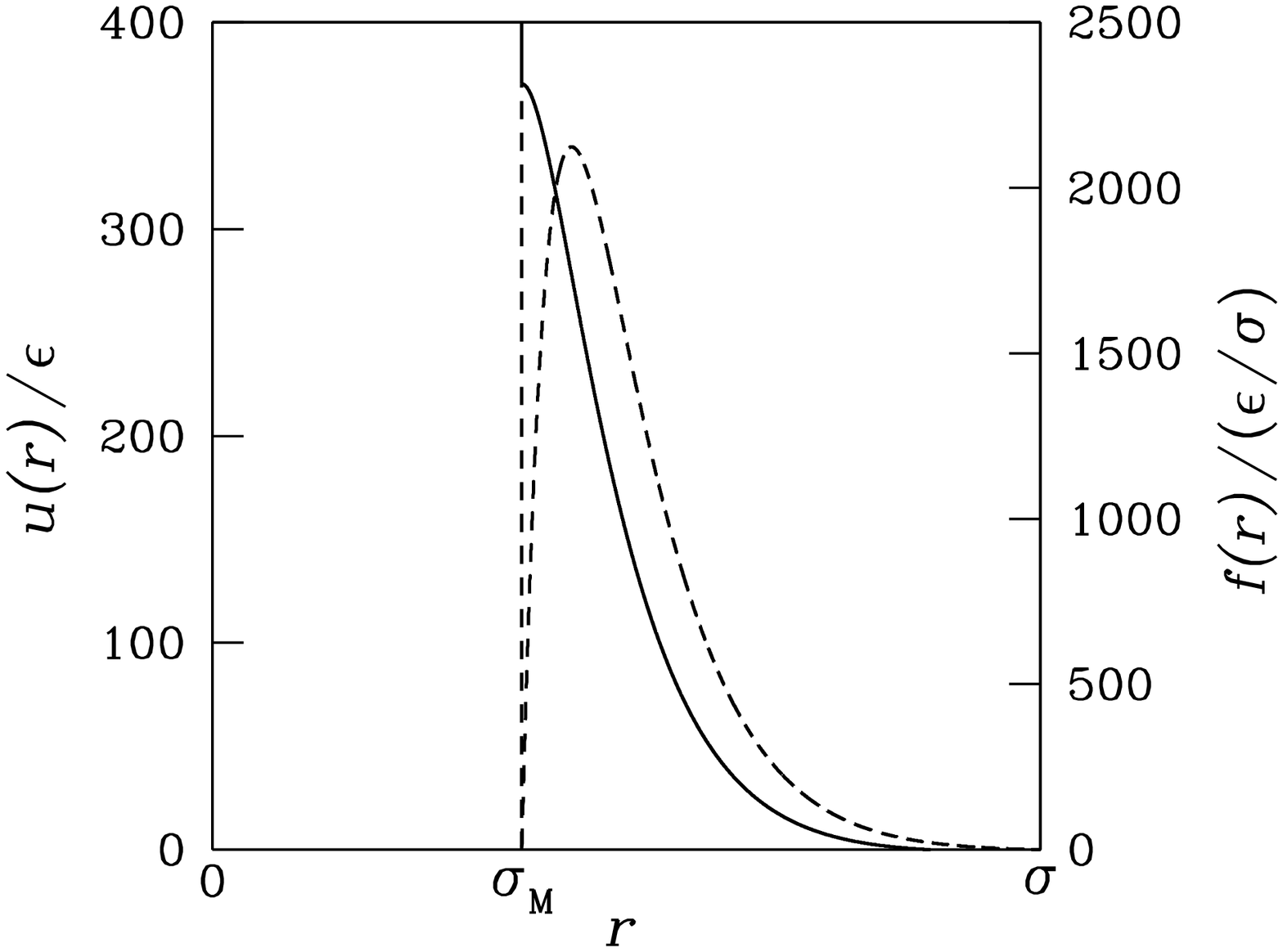}
\caption{} \label{fig1}
\end{figure}

\begin{figure}
\includegraphics[width=12cm,angle=0]{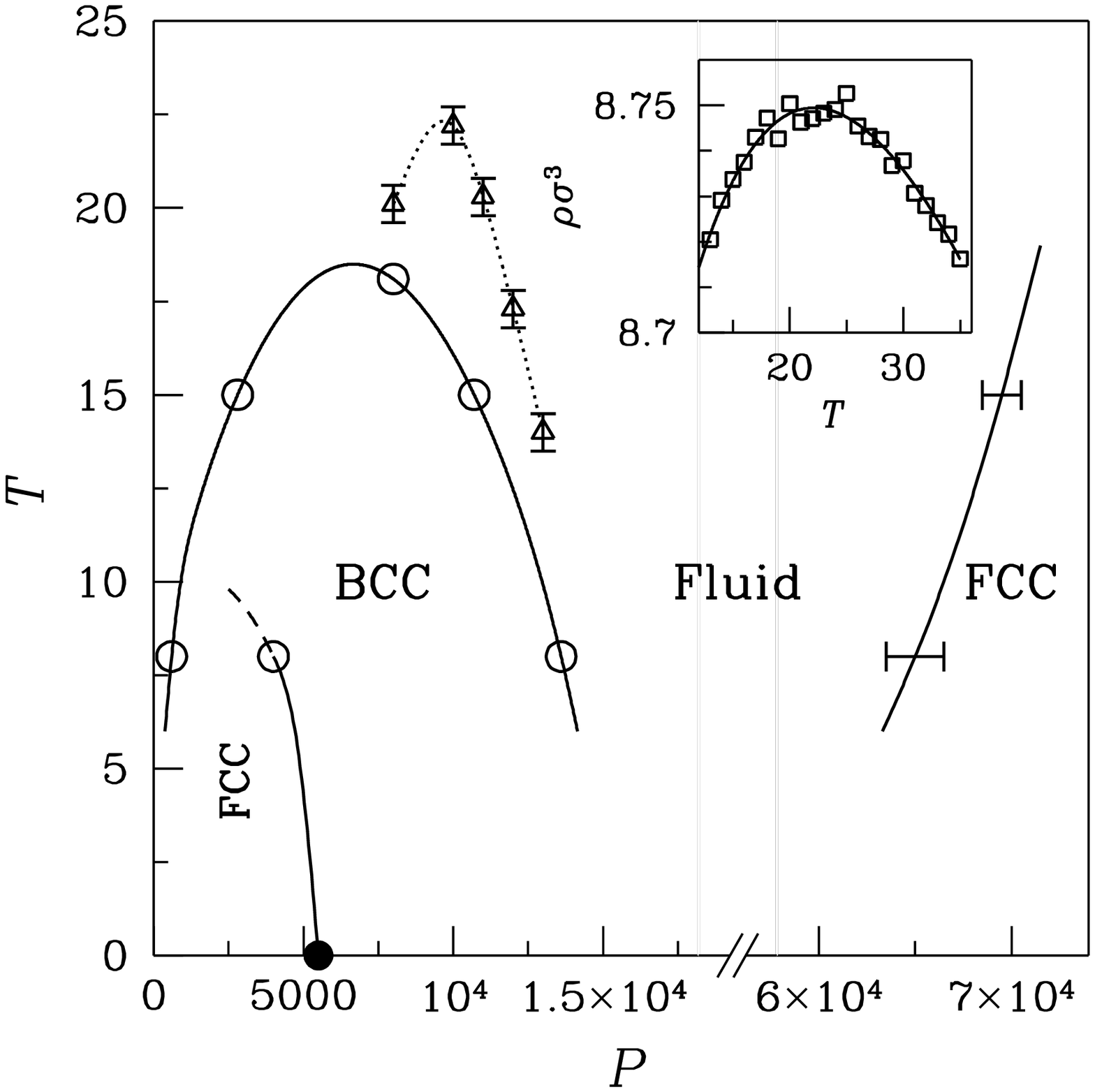}
\caption{} \label{fig2}
\end{figure}

\begin{figure}
\includegraphics[width=12cm,angle=0]{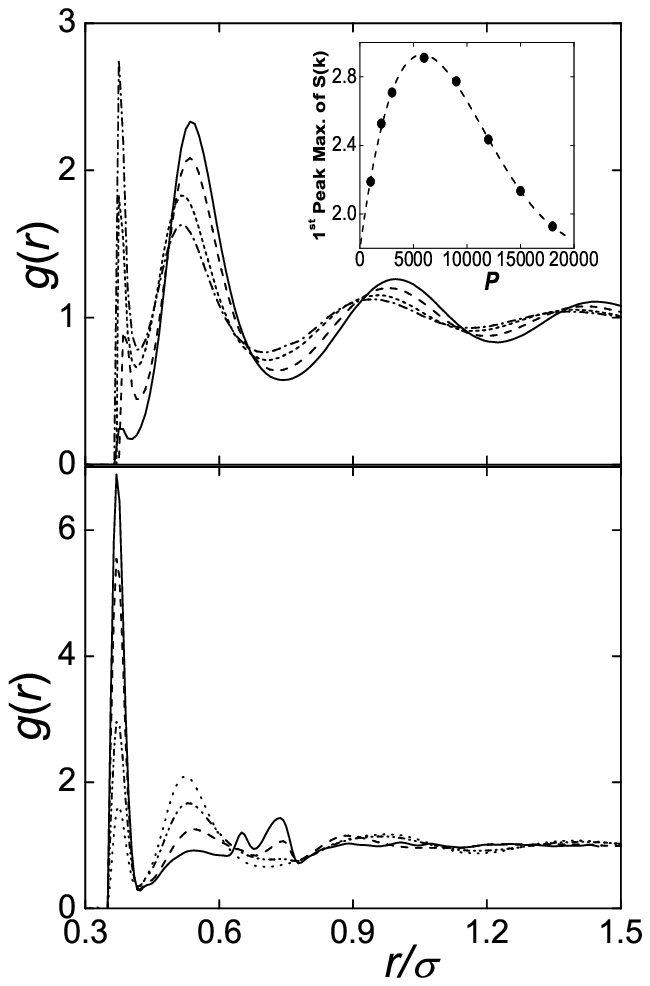}
\caption{} \label{fig3}
\end{figure}

\newpage
\begin{figure}
\includegraphics[width=12cm,angle=0]{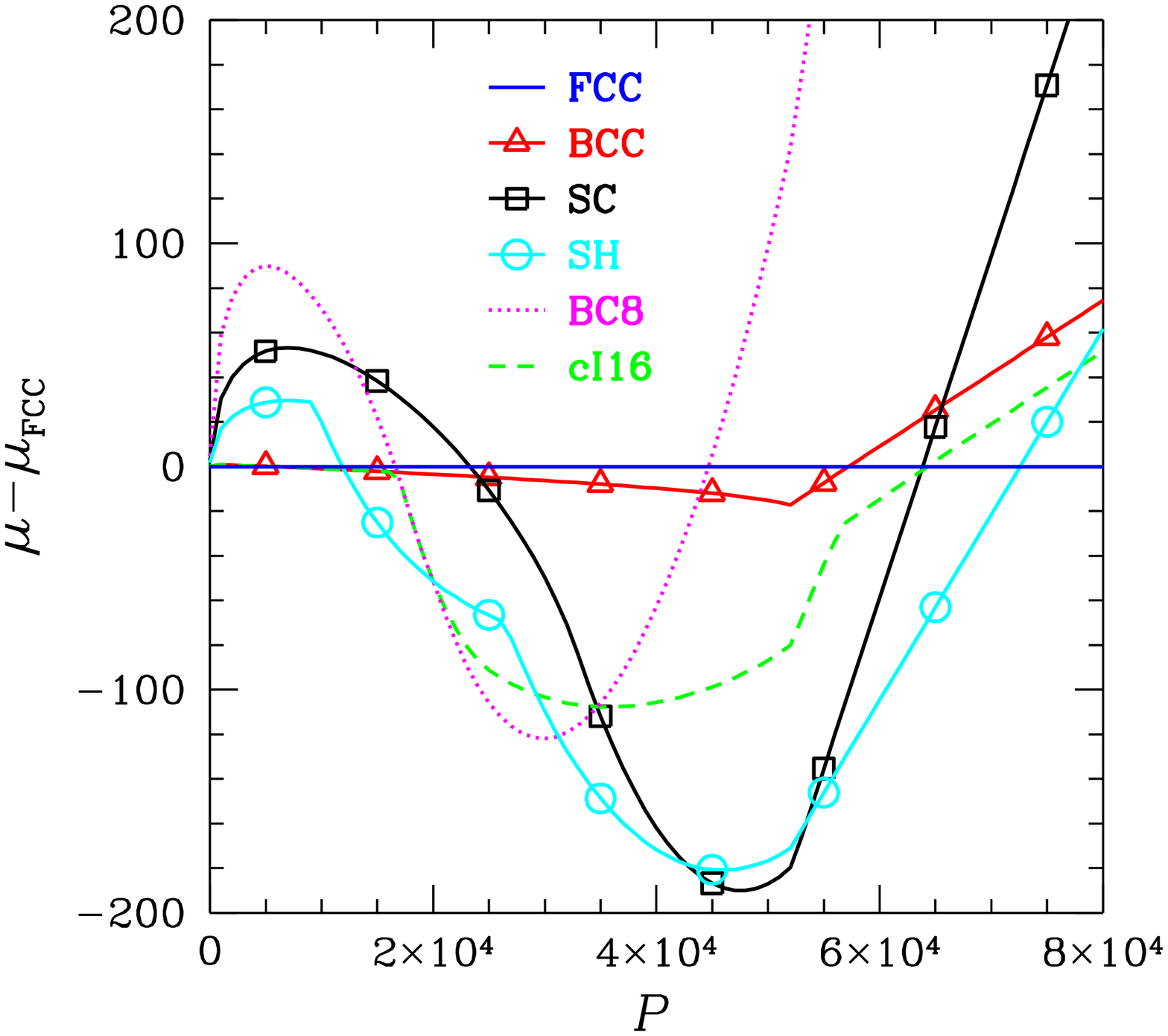}
\caption{} \label{fig4}
\end{figure}


\begin{thebibliography}{99}
\bibitem[\dag]{aff1}  E-mail: {\tt malescio@unime.it}

\bibitem[\ddag]{aff2}  E-mail: {\tt saija@me.cnr.it} (corresponding author)

\bibitem[*]{aff3}  E-mail: {\tt Santi.Prestipino@unime.it}

\bibitem{McMillan}  P. F. McMillan, {\em Nature Materials} {\bf 1},
19 (2002).

\bibitem{Young1}  D. A. Young, {\em Phase diagrams of the elements}
(University of California, Berkeley, 1991).

\bibitem{Saxena}  S. K. Saxena, G. Shen, and P. Lazor,
{\em Science} {\bf 264}, 405 (1994).

\bibitem{Boehler}  R. Boehler {\em et al.},
{\em Phys. Rev. Lett.} {\bf 86}, 5731 (2001).

\bibitem{Errandonea1}  D. Errandonea {\em et al.},
{\em Phys. Rev. B} {\bf 63}, 132104 (2001).     

\bibitem{Jayaraman}  A. Jayaraman, R. C. Newton, and J. M. McDonough,
{\em Phys. Rev.} {\bf 159}, 527 (1967).

\bibitem{Bundy}  F. P. Bundy, {\em Phys. Rev.} {\bf 115}, 274 (1959).

\bibitem{Gregoryanz}  E. Gregoryanz {\em et al.},
{\em Phys. Rev. Lett.} {\bf 94}, 185502 (2005).

\bibitem{Raty}  J-Y. Raty, E. Schwegler, and S. A. Bonev,
{\em Nature} {\bf 449}, 448 (2007).

\bibitem{Errandonea2}  D. Errandonea, R. Boehler, and M. Ross,
{\em Phys. Rev. B} {\bf 65}, 012108 (2001).

\bibitem{Rapoport}  E. Rapoport,
{\em J. Chem. Phys.} {\bf 48}, 1433 (1968).

\bibitem{Bonev}  S. A. Bonev {\em et al.},
{\em Nature} {\bf 431}, 669 (2004).

\bibitem{Mukherjee}  G. D. Mukherjee and R. Boehler,
{\em Phys. Rev. Lett.} {\bf 99}, 225701 (2007).

\bibitem{Zha}  C.-S. Zha and R. Boehler,
{\em Phys. Rev. B} {\bf 31}, 3199 (1985).

\bibitem{Royall}  C. P. Royall {\em et al.},
{\em J. Chem. Phys.} {\bf 124}, 244706 (2006).

\bibitem{Stillinger}  F. H. Stillinger, {\em J. Chem. Phys.} {\bf 65},
3968 (1976).

\bibitem{Likos}  C. N. Likos, {\em Phys. Rep.} {\bf 348}, 267 (2001).

\bibitem{Prestipino}  S. Prestipino, F. Saija, and P. V. Giaquinta,
{\em Phys. Rev. E} {\bf 71}, 050102(R) (2005).

\bibitem{Malescio1}  G. Malescio, {\em J. Phys.: Condensed Matter}
{\bf 19}, 073101 (2007).

\bibitem{Malescio2}  G. Malescio and G. Pellicane,
{\em Nature Materials} {\bf 2}, 97 (2003).

\bibitem{Stell}  P. C. Hemmer and G. Stell, {\em Phys. Rev. Lett.}
{\bf 24}, 1284 (1970).
%; G. Stell and P. C. Hemmer,
%{\em J. Chem. Phys.} {\bf 56}, 4274 (1972).

\bibitem{Young2}  D. A. Young and B. J. Alder, {\em Phys. Rev. Lett.}
{\bf 38}, 1213 (1977).
%; {\em J. Chem. Phys.} {\bf 70}, 473 (1979).

\bibitem{Jagla}  E. A. Jagla,
%{\em J. Chem. Phys.} {\bf 111}, 8980 (1999);
{\em Phys. Rev. E} {\bf 63}, 061501 (2001).

\bibitem{Debenedetti1}  P. G. Debenedetti, V. S. Raghavan, and S. S. Borick,
{\em J. Phys. Chem.} {\bf 95}, 4540 (1991).

\bibitem{Buckingham}  R. A. Buckingham,
{\em Proc. R. Soc. London, Ser. A} {\bf 168}, 264 (1938).

\bibitem{Fried}  L. E. Fried, W. M. Howard, and P. C. Souers,
{\em Exp-6: a new equation of state library for high pressure thermochemistry},
12th International Detonation Symposium, August 11-16, 2002 (San Diego, USA).

\bibitem{Ross2}  M. Ross and A. K. McMahan,
{\em Phys. Rev. B} {\bf 21}, 1658 (1980).

\bibitem{Belonoshko} A. B. Belonoshko {\em et al.},
{\em J. Chem. Phys.} {\bf 117}, 7733 (2002).

\bibitem{Saija}  F. Saija and S. Prestipino,
{\em Phys. Rev. B} {\bf 72}, 024113 (2005).

\bibitem{Frenkel}  D. Frenkel and B. Smit, {\em Understanding molecular
simulation} (Academic, New York, 2001).

\bibitem{Agrawal}  R. Agrawal and D. A. Kofke,
{\em Phys. Rev. Lett.} {\bf 74}, 122 (1995).

\bibitem{Debenedetti2}  P. G. Debenedetti, {\em Metastable liquids:
concepts and principles} (Princeton University Press, Princeton, 1996).

\bibitem{Neaton}  J. B. Neaton and N. W. Ashcroft,
{\em Nature} {\bf 400}, 117 (1999).

\bibitem{Hanfland}  M. Hanfland {\em et al.},
{\em Nature} {\bf 408}, 174 (2000).

\bibitem{McMahon}  M. I. McMahon {\em et al.},
{\em Proc. Natl. Acad. Sci. U.S.A.} {\bf 104}, 17297 (2007).

\bibitem{Tamblyn}  I. Tamblyn, J-Y. Raty, and S. A. Bonev,
{\em Phys. Rev. Lett.} {\bf 101}, 075703 (2008).

\bibitem{Pearson}  W. B. Pearson, {\em A Handbook of Lattice Spacings and
Structures of Metals and Alloys} (Pergamon, Oxford, 1967).

\bibitem{Martin}  R. M. Martin, {\em Nature} {\bf 400}, 117 (1999).

\bibitem{Ross1}  M. Ross, R. Boehler, and P. S\"oderlind,
{\em Phys. Rev. Lett.} {\bf 95}, 257801 (2005).
\end{thebibliography}
\end{document}